# Example Driven Code Review Explanation

Shadikur Rahman
Department of Computer Science
York University
North York, Canada
sadicse@yorku.ca

Umme Ayman Koana
Department of Computer Science
York University
North York, Canada
koana@yorku.ca

Maleknaz Nayebi
Department of Computer Science
York University
North York, Canada
mnayebi@yorku.ca

**Abstract**

**Background:** Code reviewing is an essential part of software development to ensure software quality. However, the abundance of review tasks and the intensity of the workload for reviewers negatively impact the quality of the reviews. The short review text is often unactionable, which needs further interaction between the reviewer and the developer. The problem becomes more critical in dynamic teams and in the case of new team members who are less familiar with their reviewers and perspectives. **Aims:** We are proposing the Example Driven Review Explanation (EDRE) method to facilitate the code review process by adding additional explanations through examples. EDRE recommends similar code reviews as examples to further explain a review and help a developer to understand the received reviews with less communication overhead. **Method:** Through an empirical study in an industrial setting and by analyzing 3,722 code reviews across three open-source projects, we compared five methods of data retrieval, text classification, and text recommendation. **Results:** EDRE using TF-IDF word embedding along with an SVM classifier can provide practical examples for each code review with 92% F-score and 90% Accuracy. **Conclusions:** The example-based explanation is an established method for assisting experts in explaining decisions. EDRE was developed based on the same philosophy and can accurately provide a set of context-specific examples to facilitate the code review process in software teams.

## 1 Introduction

Peer code review is widely regarded as a useful approach for decreasing software errors and enhancing project quality. Code review is a process where developers evaluate their code by another project developer to see whether it is of adequate quality to be incorporated into the project's core codebase [4]. The manual process and the need for intensive communication between team members to assure code quality within a peer review process is a considerable workload for a team. In modern code review and with the abundance of data around developers' activities [15, 16], tool-based review, especially automated code review tool, is becoming more popular [3]. Studies showed that a successful code review requires active participation and collaboration of reviewers and authors [20]. The manual process and the need for intensive communication between team members to assure code quality within a peer review process is a considerable workload for a team. Pascarella et al. [17] provided a taxonomy of information needs in a code review. According to this taxonomy, reviewers require information with regards to *sustainability of an alliterative solution*, *correct understanding*, *rational*, *code context*, *necessity*, *specialise expertise*, and *modularity of changes (splittable)* for a successful review process.

Analogies and examples have been human's conventional way of explaining their cognition and decisions. In today's world, the use of examples is one of the methods for decision explanation and explainable AI. Adequately explaining machine-made decisions proved to be effective in convincing domain experts to follow the recommendation and accept the intelligence of the machines. In this context, and to address the developer's need for *rational* and *code context* and to assure their *correct understanding* from the code review [3], we studied the use of analogies by providing context-specific examples along with code reviews.

The use of examples for explaining code review problem in software team have not been discussed in literature. While developers are manually searching for examples, the automated retrieval of these examples across the repositories of a software is quite possible and facilitate developer's activity. In this direction, we are answering two research questions (**RQ**s):

**RQ1:** What model can effectively identify reviews that need further explanation?
Using three major projects of an industrial partner, we labeled 3,722 reviews on a four-point Likert scale of clarity.

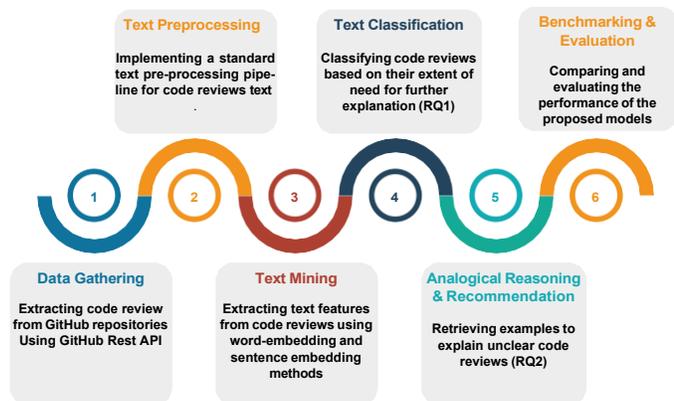

**Figure 1.** The overview of steps taken in the empirical design of this study

We then implemented and compared Logistic Regression, Multinomial Naive Bayes, SVM, Random Forest, and Gradient Boosted Regression Trees (GBRT) for classifying code review comments.

**RQ2:** What model can provide an example-based explanation for a code review?
We introduce Example Driven Review Explanation (EDRE) bot for recommending context-based examples using analogical reasoning. We used the cosine similarity on vectorized code review comments to identify analogies between reviews and generate a ranked list of examples for further explaining unclear code reviews.

Figure 1 provides an overview of the major steps taken in this study to gather, evaluate, and retrieve code review examples. In addition, we present a prototype and preliminary evaluation of EDRE as a GitHub bot that retrieves examples across the repositories of a software team upon receiving an unclear code review.

Next section provides a brief overview of the literature. Then, we describe the data used for empirical evaluation of EDRE along with the preprocessing steps we took in Section 3. Next, we discuss our study's methodology in Section 4 which involves text mining, classification, and analogical reasoning. Move to Section 5 for providing answers to RQ1 and RQ2. We further discuss the limitations of this study in Section 6.

## 2 Related Work

We briefly discuss the current state of research in code review analysis and decision explanation,

### 2.1 Code Review Analysis

Software developers spend a large amount of time and effort executing code reviews. Therefore, identifying factors that lead to useful code reviews can benefit projects by improving code review significance and quality. In a three-stage hybrid research study Bosu et al. [4], qualitatively analyzed attributes of a high quality code review. Later, in 2017 Bachelli et al. [3] identified develeoprs information need in code review process. Sadowski et al. [20] focused on Google's practice and analyzed its current state of code review. They point out in this analysis that tool-based code reviews have become the norm for a wide range of open-source and industrial systems. More recently, studies [2, 6] demonstrated that many industries have adopted modern code review as a key strategy to constantly monitor and improve the quality of code changes. Uchôa, Anderson, et al. [23] analyzed 57,498 reviewed code changes across seven open-source projects and evaluated the use of six machine learning algorithms to predict effective design changes.

Automatically recognizing the content of reviewers discussion is helpful to better understand the code review process. Li et al. [13] performed a case study on three popular open-source software projects provided on GitHub and produce a fine-grained taxonomy including 11 sub-categories for review comments. Late in 2019, Li et al. [12] standardized automated code review as a multi-instance learning task that which each transformation consisting of numerous clots is regarded as a bag, and each hunk is described as an instance. They proposed a novel deep learning model named DeepReview based on CNN. Rahman et al. [18] report a relative analysis between the useful and non-useful review comments where they determine between them using their textual features and the reviewer's experience.

While these studies have focused on improving the content of code reviews, to the best of our knowledge none have focused on explaining code review decisions.

### 2.2 Decision Explanation

Four main approaches are commonly known for explaining decisions for domain experts [1]. Code reviews and the decision to accept a change by a software developer or not is a combination of *rational decision making*, where the programming guidelines shall be followed and errors to be avoided, and *naturalistic decision making* that the expertise of the developers should be valued. Providing examples to further explain a decision is known as one of the best performed, and universal methods for explanation [10]. The rise of machine learning methods for automating decisions further extended the field of research on machine learning explanation [7]. It has been proved that humans, and in particular domain experts, follow machine-made decisions if they can reason and concur with the rational. The decision explanation in the field of software engineering is currently limited to strategic decisions such as risk management, planning, and scheduling tasks [8].

> How clear is the below code review to you?
> **"Maintain naming rules."**
> See the pull request
> ○ Unclear ○ Somewhat Unclear ○ Somewhat Clear ○ Clear

**Figure 2.** A sample of labelling task assigned to the developers. Each of the reviews were labelled by five developers.

## 3 Empirical Data

We performed an analysis on the propriety software of an industrial partner through a collaborative project. The industrial partner mainly focuses on developing mobile apps and is practicing a rigorous code review process while hosting their code on GitHub repositories. We gathered the information of three main projects using GitHub Rest API to mine the code review comments. These projects are all management tools for mobile devices. Project Alpha started in January 2019 and included 1,129 Code Reviews. Project Beta was initiated in January 2019 and included 1,767 Reviews, and The Gamma Project has been developed since 2019 and involved 826 Reviews. For all these projects, we gathered the historical information of GitHub repositories from January 2019 to November 2019, which resulted in overall 11 Months of data. These included 3,722 reviews overall and across the three projects.

We created a corpus by collecting developer code review comments (text) from GitHub projects. From here on, we refer to the code review comments simply as code review. The code review of these 3,722 reviews has, on average, 13 Words and 79 Characters in length. The relatively short length of the reviews has motivated the industrial partner to further look into ways to add clarity and explanation while not adding further workload to the reviewer.

### 3.1 Labelling Code Reviews

As the first step to answer our research questions, we labeled the code reviews. A total of 44 developers have worked directly on these three projects, including the first two authors of this paper. We launched a survey on the team management channel of the team and requested all the 44 developers to participate in labeling the data. Eight developers agreed to participate in labeling the data (response rate of 18%).

We asked the developers to evaluate the clarity of the reviews based on a four-point Likert scale of clear, Somewhat Clear, Unclear, and Somewhat Unclear. For this task, we showed the text of the code review along with a link to the conversation thread including the comments and code changes on the GitHub repository to the developer. Figure 2 shows an example of the labeling task. Each developer labeled as many code reviews as they could. Each code review have been labelled by minimum of five developers. We assigned the label of majority votes as the clarity level of the review.

As a result, we've labeled 3,722 reviews with the help of eight developers working on these projects. In Table 1, we summarized the statistics of the three projects dataset.

### 3.2 Text Prepossessing

We implemented a standard text prepossessing pipeline and applied it to all the code review comments to make it ready for our learners. This pre-processing included removing punctuation, tokenization, stop words, lemmatizing, and N-Gram extraction. We implemented these steps using `NLTK` and `BeautifulSoap` Python libraries.

Considering the short length of the reviews and the need to clean the text from unnecessary punctuation, we applied **tokenization**. We created Bag of Words from our text documents and divided each into punctuation marks making sure that short words such as "don't", "I'll", "I'd" remain as one word. Then, we used the NLTK standard stop words library for erasing the common and dispensable words from our text sentences and **removing the stop words**. To further keep and evaluate words in the context, we used **Part of Speech Tagging (POS)** to explain words in a sentence.

Then, we **lemmatized** the reviews to further transfer all the words into their lemma/root format. At this stage, by having a uniform format for data, we proceeded with **extracting N-Grams**. Using N-Grams, we maintain the sequence of candidate labels in our text document. An N-gram is a sequence of N words, which computes $p\ (w^-h)$, the probability of a word $w*$ [11]. We used the N-gram model using the user review text documents held-out data for the word-based natural language process model. We extracted bi-grams ($n$ = 2) and tri-grams ($n$ = 3) from the code reviews.

## 4 Methodology

In order to answer **RQ1** and to identify which code review needs further explanation, we performed a benchmark with five optimized Machine Learning algorithms and compared their performance. Once identifying an ambiguous code review, the EDRE bot retrieves examples (**RQ2**) across projects of the team to further explain the code review using analogical reasoning and Information Retrieval (IR) techniques.

**Table 1.** Overview of Statistics for the 3,722 labelled reviews.

| Projects | Clear | Somewhat Clear | Somewhat Unclear | Unclear |
|---|---|---|---|---|
| Alpha | 44% | 30% | 11% | 15% |
| Beta | 31% | 19% | 14% | 36% |
| Gamma | 29% | 23% | 13% | 35% |
| **Average** | 34.7% | 24.0% | 12.7% | 28.6% |

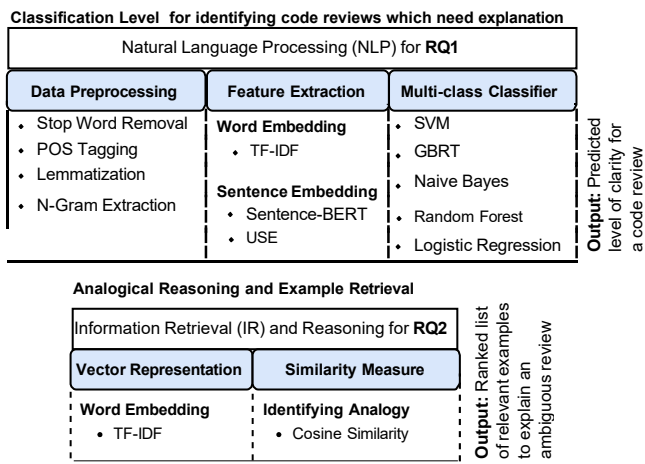

**Figure 3.** Overview of the techniques used for Example Driven code Review Explanation (EDRE)

Figure 1 shows an overview of the research. Figure 3 provides an overview of all the techniques used to answer the reserach questions and design EDRE bot.

### 4.1 Text Mining Process

To identify text features, we used the Vector Space Modeling (VSM) method by which each text is represented as vectors of identifiers. We benchmarked with both word-embedding and sentence embedding methods. Further, we evaluated the performance of three different vectorization models for code review analysis. For word embedding, We used the **TF-IDF** (Term Frequency Inverse Document Frequency) model for transforming the numerical feature vectors to make our text executable on the machine to train our models. TF-IDF is one of the most effective Information Retrieval (IR) methods for analyzing the importance of a word in a document. The TF-IDF term weighting scheme assigns a higher score to rare words and a lower one to words frequently occurring across all reviews.

Further, as the context and semantics of the reviews are quite important in analyzing the code reviews, we used sentence embedding where the semantic relatedness of each word is being considered in the review. To this end, we empirically evaluated the performance of Universal Sentence Encoder (USE) and Bidirectional Encoder Representations from Transformers (BERT) for sentence embedding in code reviews.

**Sentence-BERT** is a sentence embedding technique that uses a Siamese network architecture to enable the generation of high-quality sentence embedding. Sentence-BERT requires human annotated labeled data for fine-tuning the sentence embedding [5]. With the direct knowledge of the reviews and context, two authors of this paper provided these annotations. We also experienced with **Universal Sentence Encoder (USE)** as another sentence embedding technique [5]. USE has two separate ways of encoding sentences. The first USE model is based on a simple transformer encoder block. This model constructs an encoding subgraph of transformer architecture to generate sentence embedding. The second variant of USE also utilizes the transformer encoder attention block; however, there are some modifications done to the overall architecture compared to its first variant. The second method is called Deep Averaging Network (DAN). These two variants work together in the `TenseFlow Hub`. We used this module for implementing USE.

### 4.2 Code Review Classification

Having a vectorized representation of the 3,722 code reviews, we used popular classifiers to answer our **RQ1** as a supervised multi-class classification problem. To classify code reviews, we used supervised machine learning algorithms *Logistic Regression*, *Multinomial Naive Bayes*, *Support Vector Machine (SVM)*, *Random Forest*, and *Gradient Boosted Regression Trees (GBRT)* [11] to train our corpus datasets. To perform text classification, we used the `NLTK` library in Python. Each of the above classifiers were used on our dataset of 3,722 reviews and classifies each review either as "clear", "Somewhat clear", "Somewhat Unclear", or "Unclear".

We used cross-validation to evaluate these models. For this validation, we have used `Scikitlearn`'s `SearchGrid` function to tune our cross-validation parameter. After building the grid, we executed our `GridSearchCV` model passing `classifiers()` for finding the best estimator parameter (`.best_params_`) and n-fold value. As the result of this search, we used $n = 5$ for the cross-validation. So, among the total of 3,722 Reviews, we trained our models using 80% of the corpus (2,978 Reviews). We test the models with the remaining 20% of the data (744 Code reviews). The performance results presented from here on are the average of five times of 5-Fold cross-validation.

**Table 2.** Result of classifiers trained on various embedding models

| EMBEDDING | TF-IDF | | | | Sentence-BERT | | | | USE | | | |
|---|---|---|---|---|---|---|---|---|---|---|---|---|
| Classifiers | Precision | Recall | F-Score | Acc. | Precision | Recall | F-Score | Acc. | Precision | Recall | F-Score | Acc. |
| Random Forest | 0.94 | 0.87 | 0.90 | 0.87 | 0.88 | 0.87 | 0.86 | 0.88 | 0.88 | 0.86 | 0.87 | 0.88 |
| Naive Bayes | 0.89 | 0.88 | 0.89 | 0.87 | 0.89 | *0.89* | *0.89* | 0.89 | 0.88 | 0.89 | *0.88* | 0.88 |
| Logistic Regression | 0.92 | 0.90 | 0.91 | 0.89 | 0.87 | 0.87 | 0.88 | 0.87 | 0.88 | 0.89 | 0.86 | 0.88 |
| SVM | 0.93 | 0.91 | *0.92** | *0.90** | 0.83 | 0.82 | 0.85 | 0.83 | 0.88 | 0.87 | *0.88* | *0.89* |
| GBRT | 0.81 | 0.83 | 0.80 | 0.80 | 0.79 | 0.79 | 0.79 | 0.79 | 0.82 | 0.81 | 0.83 | 0.82 |

### 4.3 Explanation and Example Retrieval

Once a code review is somewhat or totally unclear for the developers, we form analogical reasoning to find similarities between the received code review and the formerly received and closed code reviews. Analogical reasoning is a form of knowledge management [19] and has been applied in different disciplines, including software engineering [14, 21]. Once having a code review identified as unclear or somewhat unclear, we use Cosine similarity [22] to find the top most similar reviews for the developer. Our designed model retrieves the top five most similar "clear" or "Somewhat clear" reviews.

To retrieve examples and to further clarify code reviews, we used Cosine similarity as an Information retrieval (IR) model. We computed the similarity between a received code review that needs further explanation and former code reviews. To this end, we used the cosine similarity method on vectorized code review comments. Cosine similarity measures the similarity between two vectors of an internal outcome space. It is measured by the cosine of the angle between the two vectors and determines whether the two vectors are pointing in exactly the same direction. It is often used to measure the similarity of documents in text analysis. Mathematically, it measures the cosine of angles between two vectors projected in a multidimensional space. Output values of measuring cosine similarity between two items range from zero to one, where zero means no match, and one means both items are 100% Similar.

## 5 Results & Discussion

In this section we present the results of **RQ1** and **RQ2** and further discuss the implications.

### 5.1 RQ1: Identifying Code Reviews that Need Explanation

An additional explanation for the clear decisions not only adds to the cognitive load of the experts [9]. Hence, as the first step in **RQ1** we focused on identifying the code reviews which are most in need of further explanation. We performed a comprehensive benchmark to find the best combination for the choice of embedding method and classifier as detailed in Section 4.1 and Section 4.2. In collaboration with the partnered organization and their developers, we labelled each code review as "Clear", "Somewhat clear", "somewhat unclear", and "Unclear" (See Section 3.1).

We focused on the text provided along with code reviews and extracted text features using word embedding models (here TF-IDF) and sentence embedding models (Sentence-BERT and USE). Then we trained each classifier (Random Forrest, Naive Bayes, Logistic Regression, SVM, and GBRT) with each of these three embedding models. Table 2 shows the results of classifier's performance in terms of precision, recall, F-score, and Accuracy. When using *word embedding model*, TF-IDF, SVM classifiers outperforms the other four models with f-score of 0.92 follows by Logistic regression model (with $F-score$ = 0.91) and Random Forrest (with $F-score$ = 0.90). Among all, GBRT has the worst performance in terms of both $F-score$ and $accuracy$ of 0.8.

We also implemented the Sentence-BERT encoder and USE model for *sentence embedding*. When using the Sentence-BERT model, our Naive Bayes classifier slightly outperforms other models with an $F-score$ of 0.89, while Logistic Regression comes as the second-best classifier with $F-score$ = 0.88. When it comes to the USE model, both SVM and Naive Bayes performs equally in terms of $F-score$ (= 0.88) followed by Random forest ($F-score$ = 0.87).

Overall, the difference between these models and classifiers are quite low (=0.01 difference in most cases) however, GBRT is consistently performing worst among all the models. For this reason, we further investigated the accuracy of these models as details in Table 2. Figure 4 provides a side by side comparison of F-score and accuracy of all these models. SVM trained on word embedding (TF-IDF) performs better in both metrics. Figure 5 shows the actual and predicted values for SVM model using TF-IDF word embedding in a form of a confusion matrix. The model performs very well good in identifying "clear" code reviews which could be attributed to the size of vocabulary and the Bag of Words as the three evaluated projects have been managed within the same organization.

> *SVM model trained on a simple word embedding model can best predict the level of clarity for code reviews with F-score = 0.90. The high performance of word embedding in comparison to sentence embedding could be attributed to the homogeneous use of terminology and coding practice in the analyzed project, and results might vary if looking across teams.*

### 5.2 RQ2: Explaining Code Reviews by Examples

Having identified the code review which needs further explanation, EDRE retrieves the five most similar examples of clear or somewhat clear code reviews to further explain an ambiguous review. The model is built on analogical reasoning. As a form of software knowledge management [19], we perform analogical reasoning to offer *examples* and further explain an ambiguous review. These examples were retrieved from similarity analysis and have been evaluated as clear or somewhat clear in the past. Offering these analogical examples is not meant to be prescriptive, and one example would not be understandable or suitable for all the developers.

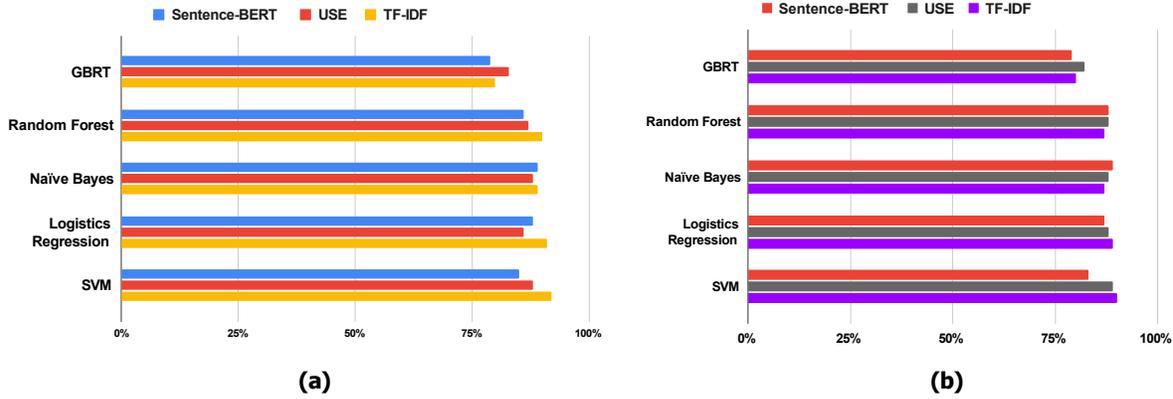

**Figure 4.** Comparison of models performance based on (a) F-score and (b) accuracy. SVM trained on word embedding (TF-IDF) performs better in both metrics.

We used Cosine similarity to find the analogy between a newly received and ambiguous code review and clear or somewhat clear examples across organizational code history. Following the idea of case-based reasoning [21], these analogical changes serve to adapt them to the specific context and to assist the developer in understanding the context and intent of the received review. EDRE tool retrieves the top five most relevant code reviews (having the highest similarity) to assist the comprehension of an ambiguous review by examples. We built a prototype explanation bot where it retrieves examples for the code reviews that is classified as "somewhat unclear" or "unclear".

Figure 6 shows the EDRE bot on GitHub. We used one-fifth of the data (744 Reviews) for testing and the rest (2,978 reviews) for the training of our EDRE tool. By that, we performed a preliminary evaluation of the retrieved examples by one of the authors who directly worked on these projects. The top five examples have been evaluated as useful or somewhat useful for explaining the received review. The full evaluation of this tool and measuring the usefulness of these examples for further explaining the code reviews remains preliminary.

> *Using Cosine similarity between code review text, EDRE can retrieve meaningful list of examples to further explain unclear code reviews. The preliminary evaluation of EDRE shows the usefulness of the retrieved examples for explaining the code review decisions.*

## 6 Limitations and Threats to Validity

Explaining code reviews by automatically retrieving examples is a new idea presented in the paper, along with a limited case study evaluation. There are several limitations attributed to the case study which should be considered when interpreting our results. While collaborating with the industry, the three chosen projects are developed within the same organization, which creates a higher probability that the size of the vocabulary would be smaller. Hence, the performance of the models can be slightly different. We hope that the detailed comparison of F-score, Accuracy, and the provided confusion matrix provides a fair baseline for the community further to evaluate the implication of our findings in their teams. We used Cosine similarity to measure textual analogy between two code reviews. This reasoning can be further extended using code review timing, reviewer's history, and other project attributes. Our ongoing study is focused on the cross-project and cross-team evaluation of EDRE using multiple attributes other than review text. It is important to note that the current evaluation of EDRE tool is limited. Further evaluation of the tool by involving more developers is needed to understand the related relevance and usefulness of the retrieved examples (**RQ2**) for developers.

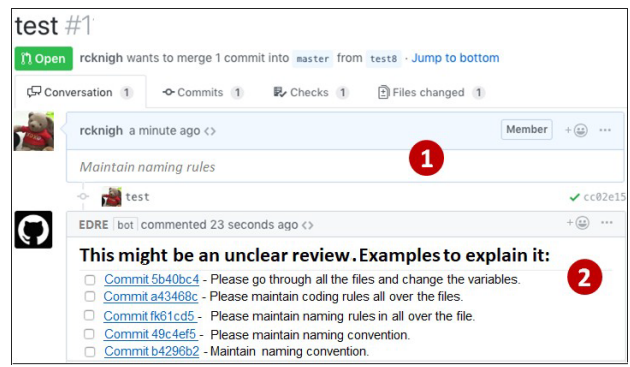

**Figure 6.** The Prototype of EDRE bot for offering examples to explain a code review. A code review is received in Step (1). Step (2) demonstrates the review is identified as "unclear" and EDRE bot provided top five most similar examples.

|  |  | **Actual label** | | | |
|---|---|---|---|---|---|
|  |  | Clear | Unclear | Somewhat clear | Somewhat Unclear |
| **Prediction label** | Clear | 638 | 11 | 18 | 1 |
|  | Unclear | 1 | 20 | 1 | 0 |
|  | Somewhat clear | 7 | 2 | 18 | 0 |
|  | Somewhat unclear | 2 | 5 | 0 | 22 |

**Figure 5.** The Confusion Matrix in Multiple labels

## 7 Conclusion and Future Work

Providing examples is one of the known forms of decision explanation. Code reviews are one of the most collaborative processes in modern software teams and are highly dependent on communications between the reviewer and developer. To reduce the communication barrier, facilitate providing feedback, and reduce review time, we designed the EDRE bot in collaboration with an industry partner. EDRE relies on an example to further explain an ambiguous code review. EDRE consists of two main parts (i) identifying ambiguous code reviews using text features and (i) retrieving a ranked list of relevant examples using analogical reasoning. In this study, we investigated 3,722 code reviews across three projects of the partnered industry. We manually labeled the level of ambiguity in each code review using software developers. We benchmarked with different embedding and vectorization models to extract textual features from code reviews and evaluated the performance of five classifiers to identify ambiguous code reviews. We further used analogical reasoning and designed the EDRE bot to retrieve examples. Our future work focuses on extending the evaluation of examples retrieved by EDRE to better understand the relevance and usefulness of the bot. In addition, we are intended to extend the empirical evaluation to investigate the performance of our proposed models across projects and organizations.

## References

[1] A Ahmad, MT Alshurideh, and BH Al Kurdi. 2021. The Four Streams of Decision Making Approaches: Brief Summary and Discussion. In *International Conference on Advanced Machine Learning Technologies and Applications*. Springer, 570–580.

[2] Eman Abdullah AlOmar, Hussein AlRubaye, Mohamed Wiem Mkaouer, Ali Ouni, and Marouane Kessentini. 2021. Refactoring practices in the context of modern code review: An industrial case study at Xerox. In *2021 IEEE/ACM 43rd International Conference on Software Engineering: Software Engineering in Practice (ICSE-SEIP)*. IEEE, 348–357.

[3] Alberto Bacchelli and Christian Bird. 2013. Expectations, outcomes, and challenges of modern code review. In *2013 35th International Conference on Software Engineering (ICSE)*. IEEE, 712–721.

[4] Amiangshu Bosu, Michaela Greiler, and Christian Bird. 2015. Characteristics of useful code reviews: An empirical study at microsoft. In *2015 IEEE/ACM 12th Working Conference on Mining Software Repositories*. IEEE, 146–156.

[5] KR1442 Chowdhary. 2020. Natural language processing. *Fundamentals of artificial intelligence* (2020), 603–649.

[6] Atacílio Cunha, Tayana Conte, and Bruno Gadelha. 2021. Code Review is just reviewing code? A qualitative study with practitioners in industry. In *Brazilian Symposium on Software Engineering*. 269–274.

[7] Leilani H Gilpin, David Bau, Ben Z Yuan, Ayesha Bajwa, Michael Specter, and Lalana Kagal. 2018. Explaining explanations: An overview of interpretability of machine learning. In *2018 IEEE 5th International Conference on data science and advanced analytics (DSAA)*. IEEE, 80–89.

[8] Jo E Hannay, Dag IK Sjoberg, and Tore Dyba. 2007. A systematic review of theory use in software engineering experiments. *IEEE transactions on Software Engineering* 33, 2 (2007), 87–107.

[9] Cengiz Kahraman, Sezi Cevik Onar, and Basar Oztaysi. 2015. Fuzzy multicriteria decision-making: a literature review. *International journal of computational intelligence systems* 8, 4 (2015), 637–666.

[10] Ralph L Keeney, Howard Raiffa, and Richard F Meyer. 1993. *Decisions with multiple objectives: preferences and value trade-offs*. Cambridge university press.

[11] Pier L Lanzi. 2000. *Learning classifier systems: from foundations to applications*. Number 1813. Springer Science & Business Media.

[12] Heng-Yi Li, Shu-Ting Shi, Ferdian Thung, Xuan Huo, Bowen Xu, Ming Li, and David Lo. 2019. Deepreview: automatic code review using deep multi-instance learning. In *Pacific-Asia Conference on Knowledge Discovery and Data Mining*. Springer, 318–330.

[13] Zhixing Li, Yue Yu, Gang Yin, Tao Wang, Qiang Fan, and Huaimin Wang. 2017. Automatic Classification of Review Comments in Pull-based Development Model.. In *SEKE*. 572–577.

[14] Maleknaz Nayebi, Homayoon Farahi, and Guenther Ruhe. 2017. Which version should be released to app store?. In *2017 ACM/IEEE International Symposium on Empirical Software Engineering and Measurement (ESEM)*. IEEE, 324–333.

[15] Maleknaz Nayebi, Homayoon Farrahi, and Guenther Ruhe. 2016. Analysis of marketed versus not-marketed mobile app releases. In *Proceedings of the 4th International Workshop on Release Engineering*. 1–4.

[16] Maleknaz Nayebi and Guenther Ruhe. 2014. An open innovation approach in support of product release decisions. In *Proceedings of the 7th International Workshop on Cooperative and Human Aspects of Software Engineering*. 64–71.

[17] Luca Pascarella, Davide Spadini, Fabio Palomba, Magiel Bruntink, and Alberto Bacchelli. 2018. Information needs in contemporary code review. *Proceedings of the ACM on Human-Computer Interaction* 2, CSCW (2018), 1–27.

[18] Mohammad Masudur Rahman, Chanchal K Roy, and Raula G Kula. 2017. Predicting usefulness of code review comments using textual features and developer experience. In *2017 IEEE/ACM 14th International Conference on Mining Software Repositories (MSR)*. IEEE, 215–226.

[19] Ioana Rus and Mikael Lindvall. 2002. Knowledge management in software engineering. *IEEE software* 19, 3 (2002), 26.

[20] Caitlin Sadowski, Emma Söderberg, Luke Church, Michal Sipko, and Alberto Bacchelli. 2018. Modern code review: a case study at google. In *Proceedings of the 40th International Conference on Software Engineering: Software Engineering in Practice*. 181–190.

[21] Martin Shepperd. 2003. Case-based reasoning and software engineering. In *Managing Software Engineering Knowledge*. Springer, 181–198.

[22] Amit Singhal et al. 2001. Modern information retrieval: A brief overview. *IEEE Data Eng. Bull.* 24, 4 (2001), 35–43.

[23] Anderson Uchôa, Caio Barbosa, Daniel Coutinho, Willian Oizumi, Wesley KG Assunçao, Silvia Regina Vergilio, Juliana Alves Pereira, Anderson Oliveira, and Alessandro Garcia. 2021. Predicting design impactful changes in modern code review: A large-scale empirical study. In *2021 IEEE/ACM 18th International Conference on Mining Software Repositories (MSR)*. IEEE, 471–482.